\documentstyle[12pt,elsart12]{article}
\newcommand{\sgommd}{$\sigma$-$\omega$ model~}
\newcommand{\CaCa}{Ca~$+$~Ca~}
\newcommand{\Elab}{$E_{\rm lab}$}
\newcommand{\vr}{\bf r}
\newcommand{\vR}{\bf R}
\newcommand{\vp}{\bf p}
\newcommand{\vP}{\bf P}

\title{\bf
Relativistic Effects in Simulations \\
of the Fragmentation Process
\\ with the Microscopic Framework
\footnote{to be published in Phys. Lett. B} }

\author{
Tomoyuki Maruyama\thanks{present address :
Advanced Science Research Center,
Japan Atomic Energy Research Institute,
Tokai, Ibaraki 319-11, Japan }
\\
Department of Physics, Kyoto University,
Kyoto 606-01, Japan \\[3ex]
Toshiki Maruyama\\
Advanced Science Research Center, \\
Japan Atomic Energy Research Institute,\\
Tokai, Ibaraki 319-11, Japan \\[3ex]
and \\[3ex]
Koji Niita\\
Advanced Science Research Center, \\
Japan Atomic Energy Research Institute,\\
Tokai, Ibaraki 319-11, Japan \\
and\\
Research Organization for Information Science and Technology, \\
Tokai, Ibaraki 319-11, Japan }
\date{}

\begin{document}

\maketitle

\begin{abstract}

We simulate the fragmentation processes in the \CaCa collisions at
the bombarding energy 1.05 GeV/u
using the Lorentz covariant RQMD and the non-covariant QMD approaches,
incorporated with
the statistical decay model.
By comparing the results of RQMD with those of QMD,
we examine the relativistic effects and
find that the multiplicity of the $\alpha$ particle after the
statistical decay process is sensitive to the relativistic effects.
It is shown that the Lorentz covariant approach is necessary to analyze the
fragmentation process even at the energy around \Elab = 1 GeV/u
as long as we are concerned with the final observables
of the mass distribution, particularly,
the light fragments around $A = 3 \sim 4$.

\end{abstract}

\vfil
\eject
\newpage

The main aim of the high and intermediate energy heavy-ion physics is
to investigate properties of nuclear matter under extreme conditions,
especially to determine the nuclear equation of state (EOS).
The numerical simulation approaches, such as BUU
\cite{Bert,BUU1} and QMD \cite{Aich}, are most promising
in describing the time-evolution of the complex system.
There are two important ingredients of transport theories:
a mean-field for nucleons and in-medium baryon-baryon
cross-sections that accounts for elastic and inelastic channels.
By searching the appropriate parameters of
the mean-field which can fit the experimental data,
one expects to know the nuclear matter properties \cite{BUU3}.

Within the framework of BUU-simulations we have succeeded to
predict/reproduce particle production data in heavy-ion collisions and
to clarify their reaction processes \cite{BUU4}.
In spite of these successes
it is not always possible to extract the nuclear EOS from
the results of theoretical calculations without ambiguities for
other model-inputs.

In order to achieve the highly dense nuclear matter, one needs to
make experiments of nucleus-nucleus collisions above about 1 GeV/u.
In such a relativistic energy region, there are important relativistic
effects other than kinematics.
The first author \cite{MARU2} has first suggested that the
transverse flow is influenced by the Lorentz contraction of the initial
phase-space distribution and the Lorentz covariance of the interaction.
In further developments the Tuebingen group have made more detailed
discussions on relativistic effects
in the dynamical evolution of the nucleus-nucleus collisions,
comparing the results of the covariant and non-covariant QMD \cite{Leh,Puri}.

As for the BUU approach, the full relativistic formalism, so called the
Relativistic BUU (RBUU) \cite{RBUU}, has been constructed by
incorporating the relativistic \sgommd \cite{Serot} with the BUU.
The first author have introduced the momentum-dependent mean-fields
into the Relativistic BUU approach and succeeded to reproduce the
experimental data of the transverse flow and the subthreshold {$K^{+}$} -
production around \Elab = 1 GeV/u simultaneously \cite{TOMO1}.
However the fragmentation has not been well discussed at this high energy,
particularly from the view point of the Lorentz covariant effects.

Recently, as a new round of the high energy heavy-ion physics,
fragment distributions are investigated most vigorously by a lot of
experimental groups \cite{expfr}.
The QMD approach is the most useful one for this study.
In the low energy regime around several tens MeV/u,
we can nicely reproduce experimental data
by simulating the statistical decay \cite{SDecay} from excited fragments
obtained at the end of the QMD calculation \cite{Toshi}.
For the high energy, however, we have the following problems
when we introduce the relativistic kinematics and the Lorentz transform
into the non-covariant QMD approach.
First, the increase of the initial density due to the Lorentz contraction
make an additional repulsion
through the density-dependent force,
which causes the spurious excitation and
the unphysical instability of initial nuclei.
It is a serious problem at the ultrarelativistic energy \cite{Sorge}.
Second we cannot correctly evaluate the internal energies of fast-moving
fragments at the end of the QMD calculation,
because the non-relativistic mean-field used in QMD
is variant under the Lorentz transformation.
Hence a fully Lorentz covariant transport approach is desired
in the relativistic energy region.

The Relativistic QMD (RQMD) approach \cite{Sorge,MARU1} must be the most
useful theoretical model for this purpose; it is formulated to describe
the interacting $N$-body system in a fully Lorentz covariant way based on
the Poincare-Invariant Constrained Hamiltonian Dynamics \cite{PICHD}.
The position and momentum coordinate, $q_i$ $p_i$, of the $i$-th nucleon
are defined as four-dimensional dynamical variables and the functions of
the time evolution parameter $\tau$.
The on-mass-shell constraints are given by
\begin{equation}
H_i \equiv p_i^2 - m_i^2 - {\tilde V_i} = 0 ,
\end{equation}
where $m_i$ and ${\tilde V_i(q_j,p_j)}$ are a mass and a Lorentz scalar
quasi potential.
The detailed form of the quasi-potential is determined by
the requirement that it is corresponding to the non-relativistic
mean-field in the low energy limit \cite{Sorge,MARU1}.
Whereas in the non-relativistic framework the argument of the potential
is a square of the relative distance between two nucleons
$\vr^2_{ij}$, in RQMD we take it as a square of the relative
distance in the rest frame of their CM system as
\begin{equation}
- q_{{\rm T}ij}^2 =
- q_{ij}^2 + \frac{ ( q_{ij} \cdot p_{ij} )^2 }{ p_{ij}^2 },
\end{equation}
\begin{eqnarray}
q_{ij} =  q_i - q_j , & p_{ij}  =  p_i + p_j .
\end{eqnarray}
The change from $\vr_{ij}^2$ to $- q_{{\rm T}ij}^2$
causes the direction dependent forces and the Lorentz contracted
phase-space distribution of fast moving matter.
In this formulation the time coordinate $q_i^0$ is distinguished from
the time evolution parameter $\tau$.
 From that one can take into account the retardation effect
approximately for the short range interaction such as usual nuclear force.
Using the above on-mass-shell and the time-fixation constraints
\cite{Sorge,MARU1}, we can describe the Lorentz covariant motions of
nucleons.
These constraints are chosen to be completely consistent
in the non-relativistic framework at the non-relativistic limit
($m_i \rightarrow \infty$).

In this paper, we calculate the fragment distribution in the \CaCa collisions
at the bombarding energy \Elab = 1.05 GeV/u with the RQMD approach plus
the statistical decay model, and then discuss the Lorentz covariant effects
in this reaction by comparing the results with those of the non-covariant QMD
approach.
In order to compare these two models precisely,
we have prepared an integrated code which includes the RQMD and QMD parts.
The common parts of this code, for example,
the prescription of the wave packets,
the initialization of the ground state,
the two-body effective interaction
and the collision prescription
are shared by the two models.
By this code we can minimize the differences
between the RQMD and QMD calculations
caused by any accidental difference in the coding.
Thus the differences observed in the final results are regarded as
an evidence of
the Lorentz covariant effects or the difference of the boosting method
discussed below.
The details of the code and the way of the actual calculations
are as in the following.

As for the basic part of the RQMD code, we follow the prescription in
Ref.~\cite{MARU1}, while for the QMD and the statistical decay codes
the basic prescriptions are taken from
Ref.~\cite{Cooling,maru0}.
For the present purpose, however, we use the relativistic kinematics
even in the non-covariant QMD approach.
Starting from the initial distributions
made with the cooling method \cite{Cooling},
we boost them according to the bombarding energy.
Then we perform the RQMD and QMD calculations and obtain the dynamical fragment
distribution.
We boost each dynamical fragment to its rest frame and evaluate
its excitation energy in the way of Ref.~\cite{maru0},
which is used in the statistical decay calculation.
We used the predictor-corrector method to integrate the equation of motion
for both models to retain strictly the energy conservation,
which is important to estimate the excitation energy of the fragments.
For the two-body effective interaction,
we use a Skyrme-type interaction with HARD EOS
(the incompressibility $K$ = 380 MeV)
and the symmetry force (the symmetry energy is 25 MeV) \cite{maru0},
while we omit the Coulomb force for  simplicity.
The widths of the wave packets are taken from the values for \CaCa
in Ref.~\cite{Toshi2}.
The Cugnon parametrization \cite{Cug} is used as a cross-section of
two baryon collisions for elastic and inelastic channels
with the frozen-delta approximation.
The delta-mass distribution and the angular distribution of
the inelastic channels are given in Ref.~\cite{GyWolf}.

As explained before, we need to transfer
the phase-space distribution twice between the rest and moving frames.
The first boost is that of the initial nuclei, namely the initial boost,
to give a bombarding energy, and the second is that of the moving fragments
at the end of the dynamical calculation into their rest frames,
namely the final boost, to evaluate their excitation energies
for the statistical decay calculation.
In the non-covariant QMD a way of these boosts is completely ambiguous;
we cannot conclude whether the boost should be of
the Lorentz or of the Galilei.
(Note that the Galilei boost does not change relative momenta and positions
of nucleons inside the nuclei and fragments in any frame.
The total momenta of the nuclei and fragments are determined by the Lorentz
transformation even in this case.)
In fact we do not have any definite way to separate a total energy of
a fragment into its translation energy and internal energy in the
non-covariant formalism using the relativistic kinematics and the
non-relativistic mean-field.
In addition the non-relativistic interaction prefers the spherical density
distribution, and then even a fast-moving nucleus or a fragment cannot
hold a Lorentz contracted shape.

\newcommand{\frag}{{\rm frag}}
\newcommand{\f}{{\rm f}}
\def\vec#1{{\bf #1}}

To examine the relativistic effects in the simulation we compare two kinds
of boosts for the non-covariant QMD calculation as follows.
One (QMD/G) is defined by the method that both the initial and
the final boosts are of Galilei.
The other way (QMD/L) is that the both boosts are of Lorentz.
The actual equations are given as follows.
For the initial boost
\begin{eqnarray}
\vr_i & = & \hat a \cdot \vr_i^0 + \vR ,
\nonumber \\
\vp_i & = & \hat b \cdot \vp_i^0 + \vP ,
\end{eqnarray}
where $\vr_i^0$ and $\vp_i^0$ are the initial position and momentum of the
$i$-th nucleon in the rest frame, and $\vR$ and $\vP$ are the mean position
and momentum of the initial nucleus, where the initial mean momentum is
evaluated from the bombarding energy with the Lorentz transformation.
In the Galilei boost the factors $\hat a$ and $\hat b$
equal to the unit matrix, while in the Lorentz boost they are given as
\begin{eqnarray}
\hat a = \hat b^{-1} &=&
1+(\gamma-1) \vec{e}_{\vec{z}} {}^{\rm t}\vec{e}_{\vec{z}}
= \pmatrix{
1&0&0\cr
0&1&0\cr
0&0&\gamma\cr},
\nonumber\\
\gamma &=& \frac{1}{\sqrt{1 - v^2}}
\nonumber\\
\end{eqnarray}
with the boosting velocity $v$ and the unit vector
$\vec{e}_{\vec{z}}$ along the
beam direction.
For the final boost
\begin{eqnarray}
\vr_i^{\frag} & = & \hat a_\f \cdot (\vr_i - \vR^{\frag}) ,
\nonumber \\
\vp_i^{\frag} & = & \hat b_\f \cdot (\vp_i - \vP^{\frag}) ,
\end{eqnarray}
where $\vR^{\frag}$ and $\vP^{\frag}$ are given as
\begin{eqnarray}
\vR^{\frag} & = & \sum_{i \in \frag} {\vr}_i / A_\frag,
\nonumber \\
\vP^{\frag} & = & \sum_{i \in \frag} {\vp}_i / A_\frag.
\end{eqnarray}
Here the factors $\hat a_\f$ and $\hat b_\f$ equal to the unit matrix
in the Galilei boost, whereas they are given in the Lorentz boost as
\begin{eqnarray}
\hat a_\f = \hat b_\f^{-1} &=&
1+(\gamma_\f-1)\vec{e}_{v_\f}{}^{\rm t}\vec{e}_{v_\f},
\nonumber \\
\gamma_\f &=& \frac{1}{\sqrt{1 - v_\f^2}}
\nonumber \\
\end{eqnarray}
with the fragment velocity $v_\f$ and the unit vector
$\vec{e}_{v_\f}$ along the fragment velocity.

In RQMD the definition of the fragments produced in the dynamical process
is not trivial due to the multi-time description.
We thus approximately define
the fragments at the end of the RQMD simulation in the following way.
First, we search candidates of the dynamical fragments, where all
nucleons have neighbours with the condition that $- q_{ij}^2 < 16$ fm$^2$.
We then boost the nucleons inside each candidate into its rest frame
by the Lorentz transformation.
With the above condition, in addition to
the cluster separability condition employed in RQMD \cite{Sorge},
we can roughly define the cluster in which the time competent of $q_i$ of
all nucleons are not so much different each other.
Finally we again search dynamical fragments in this frame
with the same procedure used
in the non-covariant QMD \cite{Toshi2}.
We note that
the final result of the mass distribution of the fragments obtained
with this method does not change within the statistical errors,
if we vary the value of the above condition
from 14 fm$^2$ up to 18 fm$^2$.
We also apply this method to the non-covariant QMD
with the two kinds of the boosts mentioned above.

Before studying relativistic effects, we check
the low energy limit where RQMD and QMD should give the same result.
We show in Fig.~1 the impact-parameter dependence of the
$\alpha$ multiplicity in \CaCa collisions at the bombarding energy
\Elab = 50 MeV/u.
This figure surely shows that
the results of RQMD agree with those of QMD at the low energy.

Next we simulate \CaCa collisions at the bombarding energy \Elab = 1.05 GeV/u.
In the left-hand side of Fig.~2 we show the impact-parameter dependence
of the dynamical single proton (a) and the $\alpha$ (b) multiplicities
obtained in RQMD, QMD/G and QMD/L calculations before the statistical decay.
It can be easily seen that there is no clear difference between
the results of these calculations.
Though the Lorentz contraction in QMD/L causes the unphysical instability
in the initial distribution as mentioned before, the figure shows that
this instability is not so significant as to emit nucleons
in the dynamical stage at this energy around \Elab = 1 GeV/u.

In the right-hand side of Fig.~2, on the other hand,
we show the same quantity as in the left-hand side but
after the statistical decay.
One can see obvious differences in the results of these models.
The multiplicity of single protons (Fig.~2c) decreases
with increasing the impact parameter $b$ and becomes
almost zero at $b \approx 8$ fm in
RQMD and QMD/G,
while it has still a finite value in QMD/L.
In such peripheral region two-baryon collisions rarely occur
so that this result of the QMD/L simulations is unphysical.
This unphysical behavior is much enhanced
in the $\alpha$ multiplicity (Fig.~2d).
These are due to the spurious excitation in the QMD/L calculation
caused by the initial Lorentz boost
and the non-covariant treatment of the interaction.
The QMD/G approach seems to be free, at least, from this problem.

In Fig.~2d the multiplicity of the $\alpha$ particle is shown as its
impact-parameter dependence.
By comparing the results between before (b) and after (d)
the statistical decay process,
it is clear that the $\alpha$ particles are produced dominantly
in the evaporation process from the excited fragments.
These excited fragments are coming from the spectator zone,
since the participant breaks mainly into single
nucleons in this energy region.
Thus the $\alpha$ multiplicity is large in the middle impact parameter
region, as seen in Fig.~2d.
In this middle impact parameter region, around $b$ = 4 fm,
we can see the clear difference between the results of RQMD and
QMD/G, which amounts to roughly 50 \%.

In  RQMD,
the Lorentz contracted initial distribution compacts the size of the
participant zone and increases the maximum density \cite{MARU1} and
nucleon momenta in this zone.
Then the participant in RQMD is more heated \cite{Momd},
and makes a stronger repulsion \cite{MARU2,Leh} than in QMD/G.
Furthermore, the reaction time is shorter in RQMD \cite{MARU1},
and the spectators and  participant are separated more clearly
in the momentum space
in RQMD than in QMD \cite{Puri}.
As a result, the average temperature of the spectators is
lower in RQMD than in QMD/G.
This difference between RQMD and QMD/G also affects
the proton multiplicity in Fig.~2c.
However, the proton is mainly produced in the dynamical process
(cf. Fig.~2 a and c).
Thus this relativistic effects appear only in the
$\alpha$ multiplicity.

Finally, we show the total cross-section of fragments with
the mass number $A$ after the statistical decay in Fig.~3.
Around $A$ = 3 and 4, which are mainly evaporated from the excited spectators,
we can see the clear difference about factor of two between the results of RQMD
and QMD/G.
It is noted that before the statistical decay the mass distributions
of the fragments are almost the same in RQMD and QMD/G within the statistical
errors.
Therefore, the difference around $A$ = 3 and 4 is attributed to
the difference of the internal energies of
the fragments produced by these models.

In summary, we calculate the fragment distribution using the RQMD and
QMD approaches combined with the statistical decay model.
The mass distributions of the fragments are almost equivalent in the
dynamical stage between RQMD and QMD at the energy around 1 GeV/u.
This agrees with the conclusion of Ref.~\cite{Puri},
where it is claimed that the relativistic effects on the observables
like the resonance production, density and rapidity distributions
are less significant at the energy \Elab $\le$ 2 GeV/u.
However, the difference appears in the internal energies of the
fragments produced by these models.
We have shown that the multiplicity of the $\alpha$ particle after the
statistical decay process is very sensitive to this difference of the
internal energies of the fragments.
Naive extension of the non-covariant QMD with the Lorentz contraction
of the initial boost, called QMD/L in this paper,
causes the unphysical instability and the spurious
excitation of the fragments.
This leads to an unphysical result of the $\alpha$ multiplicity even
at \Elab = 1 GeV/u.
Though the QMD with the Galilei boost (QMD/G) is almost free from the spurious
excitation of the fragments,
the relativistic effects in the dynamical process change the internal
energies of the fragments and thus clearly appear in the final $\alpha$
multiplicity as a factor of two difference compared with the QMD/G.
Therefore we have to use the Lorentz covariant approach to analyze the
fragmentation process even at the energy around \Elab = 1 GeV/u
as long as we are concerned with the final observables
of the mass distribution, particularly
the light fragments around $A = 3 \sim 4$.

In the actual analysis of the fragmentation process
by the QMD and/or RQMD in comparison with the data,
many other ingredients of the models are responsible to the final
mass distributions.
For example, the explicit momentum-dependent force
has been already indicated to have a very important role
in this energy region \cite{TOMO1,Momd}.
The Coulomb force is also supposed to influence
the fragmentation process, particularly the expansion phase
of the hot and/or dense matter.
These are not considered in this work.
Although it is not clear whether
the relativistic effects discussed above are always
visible in the final mass distribution or not
when we include these ingredients in the models,
this relativistic problem always exists in a fundamental point
of the formulation.
In order to clarify the effects of these other ingredients,
it is also necessary to use the Lorentz covariant approach
at least above \Elab = 1 GeV/u.

\bigskip

We would like to thank Prof. H. Horiuchi  for the useful
discussions.
This work is financially supported in part by the RCNP, Osaka
University, as a RCNP Computational Nuclear Physics Project
(Project No~94-B-04).
Tomoyuki Maruyama is also supported by JSPS fellowship (No.~2207).



\newpage

\noindent
{\large\bf Figure captions}\hfill
\vspace{1em}

\noindent
{\bf Fig.~1:} The impact-parameter dependence of the multiplicities of
the $\alpha$ particle in \CaCa collisions
at \Elab = 50 MeV/u.
The full squares and open circles show the results of the
RQMD and QMD/G  approaches, respectively.

\vspace{1em}\noindent
{\bf Fig.~2:}
The impact-parameter dependence of the multiplicities of
the proton (upper) and $\alpha$ particle (lower) in \CaCa collisions
at \Elab = 1.05 GeV/u before the statistical decay (left-hand side)
and after that (right-hand side).
The full squares, open circles, and diamonds show the results of the
RQMD, QMD/G, and QMD/L, respectively.

\vspace{1em}\noindent
{\bf Fig.~3:} The total cross-section of fragments with the
mass-number $A$.
The full squares and open circles show the results of the
RQMD and QMD/G  approaches, respectively.

\eject

\begin{thebibliography}{99}

\bibitem{Bert} G. F. Bertsch and S. Das Gupta, Phys. Rep. {\bf 160} (1988)
189.
\bibitem{BUU1} W. Cassing and U. Mosel, Prog. Part. Nucl. Phys. {\bf 25}
(1990) 235.
\bibitem{Aich} J. Aichelin, Phys. Rep. {\bf 202} (1991) 233, and
reference therein.
\bibitem{BUU3} For example,
H. Stoecker and W. Greiner, Phys. Rep. {\bf 137} (1986) 277.
\bibitem{BUU4} W. Cassing, W. Metag, U. Mosel and K. Niita, Phys.
Rep. {\bf 188} (1990) 363.
\bibitem{MARU2} T. Maruyama, G.Q. Li and A. Faessler,
Phys. Lett. {\bf 268B} (1991) 160.
\bibitem{Leh} E. Lehmann, R.K. Puri, A. Faessler, T. Maruyama,
G.Q .LiAN. Ohtsuka, S.W. Huang, D.T.KhoaAY.LotfyAM. A. Matin,
Prog. Part. Nucl. Phys. 30 (1992) 219;\\
E. Lehmann, R. K. Puri, A. Faessler, G. Batko and S.W. Huang,
Phys. Rev. {\bf C} in press.
\bibitem{Puri} R. K. Puri, E. Lehmann, A. Faessler and S.W. Huang,
Z. Phys. {\bf A351} (1995) 59.
\bibitem{RBUU} C.M. Ko, Q. Li and R. Wang, Phys. Rev. Lett. {\bf 59}
(1987) 1084;\\
B. Blaettel, V. Koch and U. Mosel, Rep. Prog. Phys. 56 (1993) 1.
\bibitem{Serot} B.D. Serot and J. D. Walecka, The relativistic Nuclear
Many Body Problem.  In J. W. Negele and E. Vogt, editors,
$Adv. Nucl. Phys. {\bf Vol. 16}$, page 1, Plenum Press, 1986,
and reference therein.
\bibitem{TOMO1} T. Maruyama, W. Cassing, U. Mosel, S. Teis and K. Weber,
Nucl. Phys {\bf A552} (1994) 571.
\bibitem{expfr} For example,
L. G. Moretto and G. J. Wozniak, Ann. Rev. Nucl. Part. Science
{\bf 43} (1993) 379;\\
A. S. Botvina et al., GSI-94-36 Preprint and references therein.
\bibitem{SDecay} F. Puehlhofer,  Nucl. Phys. {\bf A280} (1977) 267.
\bibitem{Toshi} T. Maruyama, A. Ono, A. Ohnishi and H. Horiuchi,
Prog. Theor. Phys. {\bf 87} (1992) 1367.
\bibitem{Sorge} H. Sorge, H. Stoecker and W. Greiner, Ann. of
Phys. {\bf 192} (1989) 266.
\bibitem{MARU1} T. Maruyama, S.W. Huang, N. Ohtsuka, G.Q. Li, A. Faessler
and J. Aichelin, Nucl. Phys. {\bf A534} (1991) 720.
\bibitem{PICHD} For Example, P. A. M. Dirac, Rev. Mod. Phys.
{\bf 21} (1949) 392;\\
A. Komer, Phys. Rev. {\bf D18} (1978) 1881, 1887, 3617;\\
J. Samuel, Phys. Rev. {\bf D26} (1982) 3475, 3482.
\bibitem{Cooling} T. Maruyama, A. Ohnishi and H. Horiuchi, Phys. Rev
{\bf C42} (1990) 386.
\bibitem{maru0} T. Maruyama, A. Ohnishi and H. Horiuchi, Prog. Theor. Phys.
{\bf 87} (1992) 1367.
\bibitem{Toshi2} T. Maruyama, A. Ohnishi and H. Horiuchi,
Phys. Rev. {\bf C45} (1992) 2335.
\bibitem{Cug} J. Cugnon, T. Mizutani, J. Vandermeulen, Nucl. Phys.
{\bf A352} (1981) 505.
\bibitem{GyWolf} Gy. Wolf, G. Batko, W. Cassing, U. Mosel,
K. Niita and M. Shaeffer, Nucl. Phys. {\bf A517} (1990) 615
\bibitem{Momd}
G.Q. Li, S.W. Huang, T. Maruyama, D.T. Khoa, Y. Lotfy and A. Faessler,
Nucl. Phys. {\bf A537} (1992) 631;\\
G.Q. Li, T. Maruyama, Y. Lotfy, S.W. Huang, D.T. Khoa and A. Faessler,
Nucl. Phys. {\bf A537} (1992) 645.

\end{thebibliography}
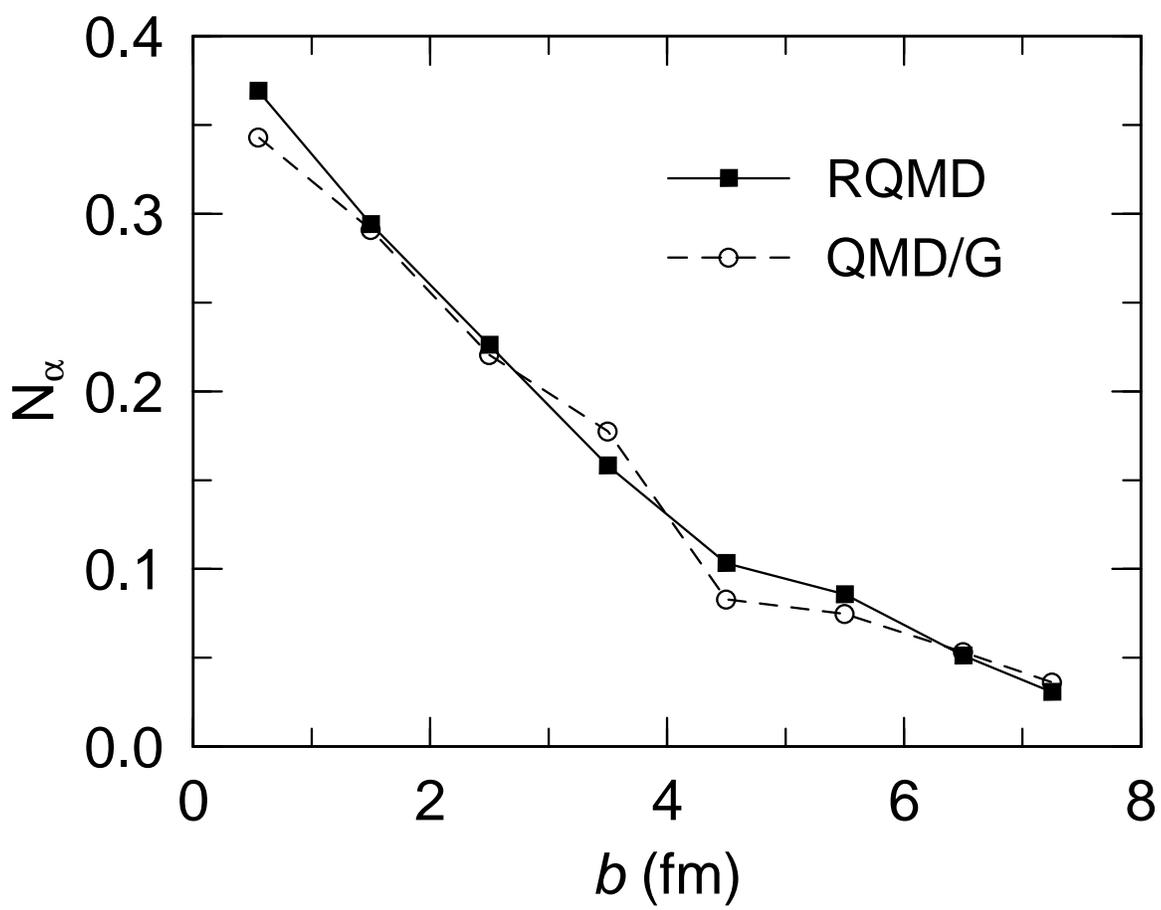
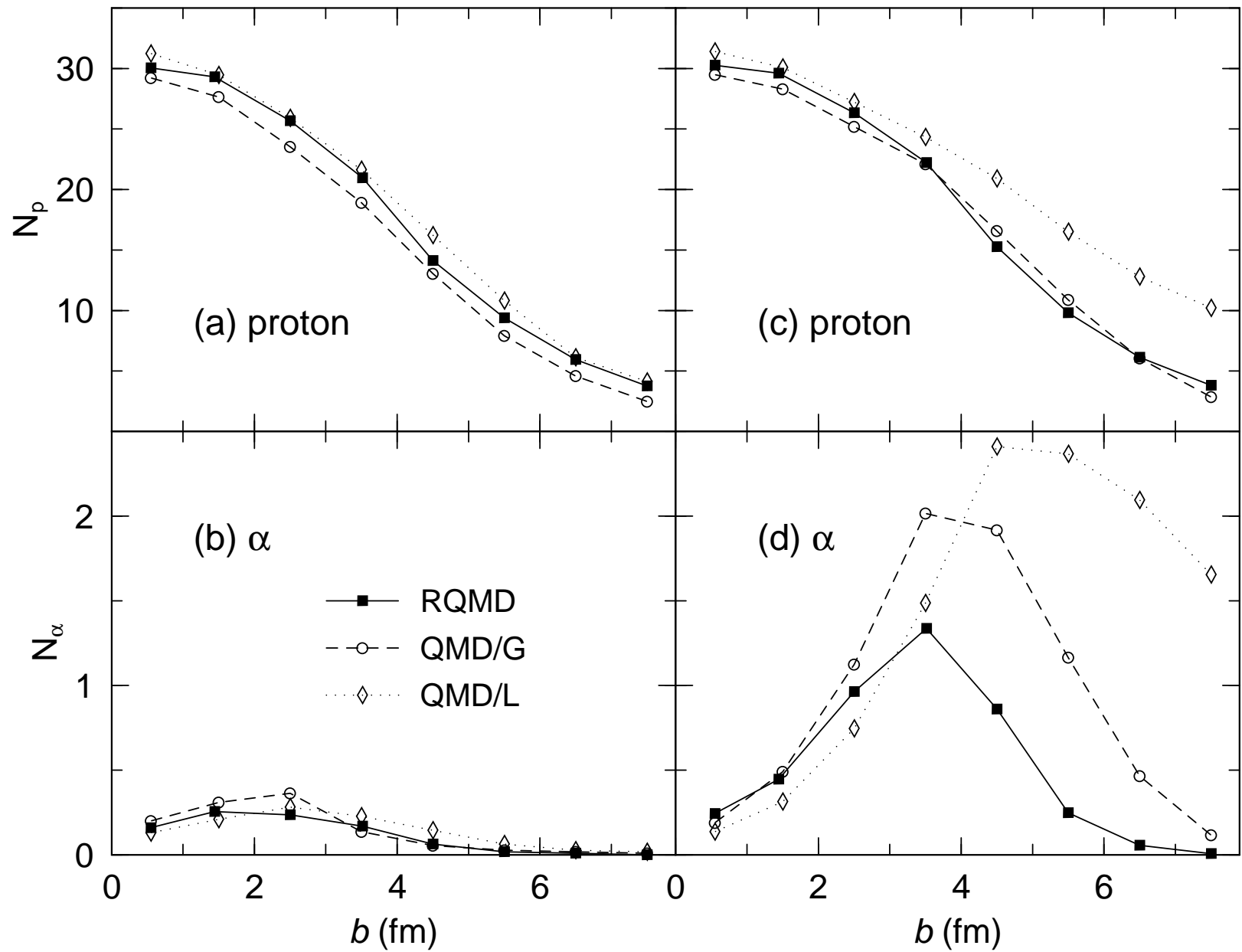
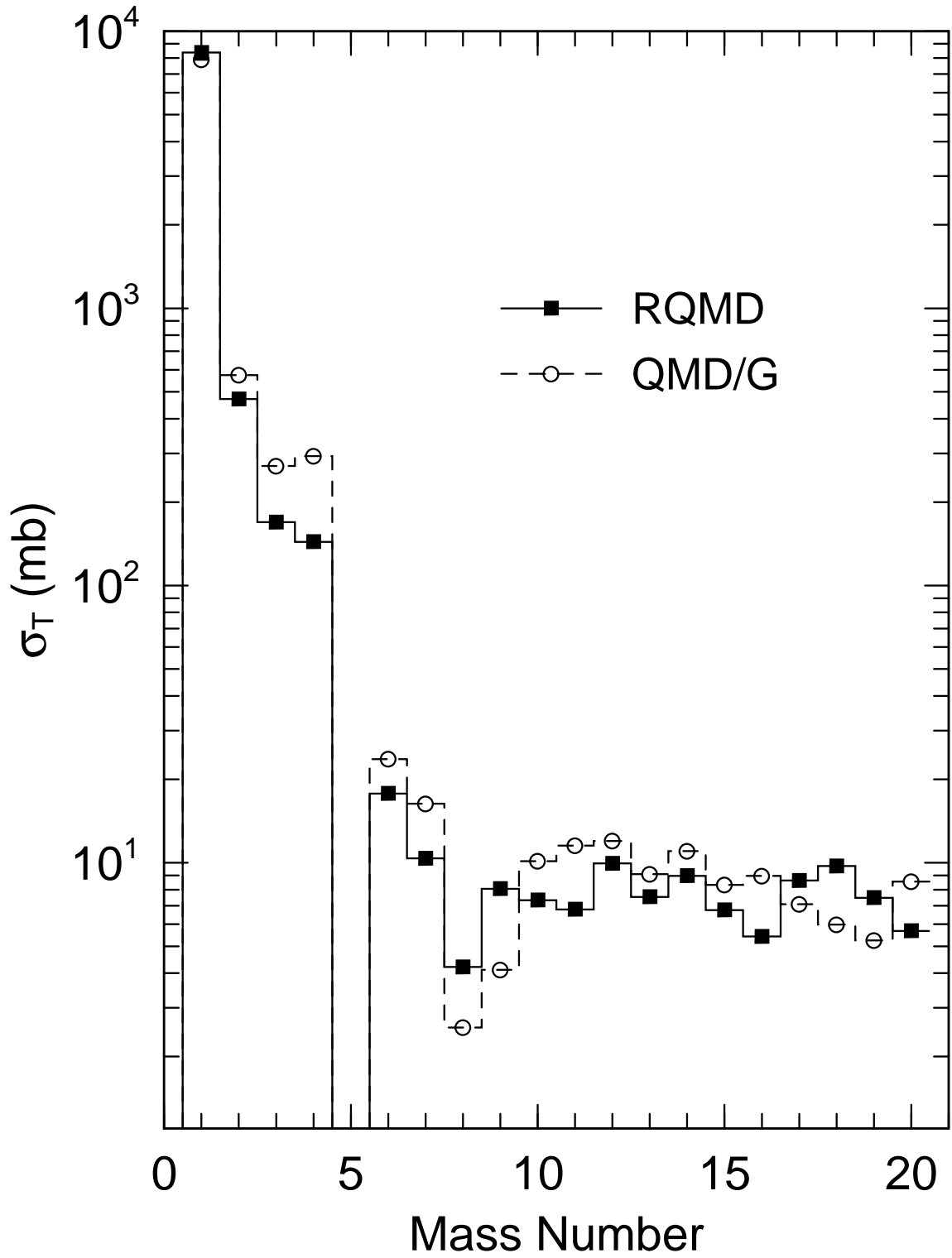
\end{document}